\documentclass[letterpaper,twocolumn,10pt]{article}
\usepackage{hyperref}
\usepackage{usenix-2020-09}
\usepackage{booktabs}
\usepackage{tikz}
\usepackage{amsmath,amssymb}
\usepackage{mathtools}
\usepackage{multirow}
\usepackage{rotating}
\usepackage{threeparttable}
\usepackage{filecontents}
\usepackage{xspace}
\usepackage{array}
\usepackage[group-separator={,}]{siunitx}
\usepackage{caption}
\usepackage{subcaption}

\newcolumntype{H}{>{\setbox0=\hbox\bgroup}c<{\egroup}@{}}

\newcommand{\tool}{Balboa\xspace}

\definecolor{mymagenta}{HTML}{741b47}
\definecolor{mygreen}{HTML}{38761d}
\definecolor{myyellow}{HTML}{fff2cc}

\definecolor{mygreen2}{HTML}{66c2a5}
\definecolor{myorange2}{HTML}{fc8d62}
\definecolor{mypurple2}{HTML}{8da0cb}

\def\Snospace~{\S{}}

\newcommand{\KDF}{\ensuremath{\mathrm{KDF}}\xspace}

\begin{document}
\date{}
\title{
    Balboa:
    Bobbing and Weaving around Network Censorship
}
\author{
{\rm Marc B. Rosen}\\
Galois, Inc.
\and
{\rm James Parker}\\
Galois, Inc.
\and
{\rm Alex J.\ Malozemoff}\\
Galois, Inc.
}

\maketitle

\begin{abstract}
    We introduce \tool, a link obfuscation framework for censorship
    circumvention. \tool provides a general framework for tunneling data through
    existing applications. \tool sits between an application and the operating system,
    intercepting outgoing network traffic and rewriting it to embed data. To
    avoid introducing any distinguishable divergence from the expected
    application behavior, \tool only rewrites traffic that matches an externally
    specified \emph{traffic model} pre-shared between the communicating parties.
    The traffic model captures some subset of the network traffic (e.g., some
    subset of music an audio streaming server streams). The sender uses this
    model to replace outgoing data with a pointer to the associated location in
    the model and embed data in the freed up space. The receiver then extracts
    the data, replacing the pointer with the original data from the model before
    passing the data on to the application. When using TLS, this approach means
    that application behavior with \tool is \emph{equivalent}, modulo small
    (protocol-dependent) timing differences, to if the application was running
    without \tool.

    \tool differs from prior approaches in that it (1) provides a framework for
    tunneling data through arbitrary (TLS-protected) protocols/applications, and
    (2) runs the unaltered application binaries on standard inputs, as opposed to most prior
    tunneling approaches which run the application on non-standard---and thus
    potentially distinguishable---inputs.

    We present two instantiations of \tool---one for audio streaming and one for
    web browsing---and demonstrate the difficulty of identifying \tool by a
    machine learning classifier.
\end{abstract}
\section{Introduction}
\label{sec:introduction}

The continued increase in Internet censorship across the
world~\cite{FreedomOnTheNet} has spurred the research community to develop
censorship resistant systems (CRSs). These systems seek to allow a party within
a monitored region to access censored content. In this work we focus
specifically on CRSs based on \emph{link obfuscation}. Link obfuscation aims to
allow communication between two or more parties such that a censor monitoring
(or manipulating) the network should not be able to detect such communication.
There are a wide array of such tools (see Khattak et al.'s
systemization~\cite{Khattak2016a} for a detailed summary of CRSs---including
those that focus on link obfuscation---as of 2016) but they tend to fall into
two main categories: \emph{look-like-nothing} approaches, which avoid detection
by being hard to classify as any particular type of traffic, and
\emph{look-like-something} approaches, which generate traffic designed to look
like a protocol the censor does not wish to block. Look-like-something
approaches, themselves, generally fall within two camps: \emph{mimicry} and
\emph{tunneling}.

In the \emph{mimicry} approach, a CRS produces network traffic designed to be
close to what an existing implementation of the target protocol might look like.
One difficulty with this approach is that it is nearly impossible for the CRS to
perfectly mimic an existing implementation of the target protocol. Any
differences between this implementation and the CRS are features that a
sufficiently powerful censor could target. In practice, CRSs that take the
mimicry approach tend to have easily distinguishable features, leading
Houmansadr et al.~\cite{Houmansadr2013a} to argue that mimicry approaches are
``fundamentally flawed.''

An alternative approach called \emph{tunneling} directly runs a concrete
implementation of the target protocol, addressing the key concern of the
mimicry approach.
Unfortunately, the tunneling approach is also imperfect due to the use of
\emph{non-standard} input to the concrete implementation. As an example,
DeltaShaper~\cite{Barradas2017a} is a CRS that tunnels user data through Skype
calls by encoding user data as audio and video data which is then fed into Skype via
a simulated camera and microphone. The receiving party extracts the data by
processing the call's output. Even though Skype data is encrypted, Wright et
al.~\cite{SkypeEncryptionIsNotEnough} found that the sizes and timings of
packets alone can still leak information about the plaintext. As a result, a
censor who can observe the encrypted packets can determine that the inputs to
the Skype call are not standard inputs (e.g., the audio sounds like a dial-up
modem instead of somebody talking). While Barradas et al.~\cite{Barradas2017a}
implemented techniques in DeltaShaper to try to mitigate this information leak,
the same authors later showed~\cite{Barradas2018a} that the mitigation was
insufficient, and that (given labelled training data) a censor could discover
when DeltaShaper was in use.

\medskip

In summary, mimicry approaches can be detected because a CRS is unlikely to
perfectly match a concrete implementation of the target protocol, and tunneling
approaches can be detected because the concrete implementation of the target
protocol is not run on standard inputs.

\subsection{Our Approach}\label{sec:our-approach}

In this work, we introduce \tool, a link obfuscation framework that aims to
address the above concerns by running a \emph{concrete application} implementing the
target protocol on \emph{standard inputs}. The key insight is that if the
communicating parties know \emph{a priori} some subset of the expected network
traffic then that network traffic does not actually need to be sent, and could
instead be replaced by arbitrary data. \tool handles this by sitting between the
concrete application and operating system, intercepting outgoing and incoming network data. In
addition, the communicating parties have a pre-shared \emph{traffic model} which
contains some subset of the expected network traffic. Whenever \tool on the
sender side intercepts outgoing data contained in the model, it replaces said
data with a \emph{pointer} to the appropriate location in the model; \tool on
the receiver side then ``inverts'' this procedure by using its own model to
replace the pointer with the actual data.

This approach has two key features: (1) the applications themselves act
\emph{exactly the same} as if \tool were not running, and (2) the sender can
insert arbitrary data into the ``freed up'' bytes, since the pointer is
much smaller than the data that would have been sent. Importantly, \tool \emph{does
  not} assume that the traffic model is complete (or even accurate). Instead,
\tool first checks to see whether outgoing traffic matches the traffic model
before performing rewrite operations. If part of the outgoing traffic does not
match the model, \tool does not modify it.

\tool relies on TLS to hide the fact that the application data itself
changed---all other (non-timing) characteristics of the traffic (e.g., TLS
record length) remain identical. In particular, \tool uses debugging features
found in most TLS libraries to extract the session key and uses this to decrypt
and re-encrypt the intercepted TLS traffic.

Because \tool only makes changes to the plaintext content of TLS-protected
network traffic, the fact that \tool is running is indistinguishable to a censor
lacking the session key for the connection, modulo a small protocol-dependent
timing delay (cf.\ \autoref{sec:evaluation}). Importantly, unlike many
censorship circumvention approaches, \tool does not modify the TLS handshake at
all. This makes it much more difficult for the many censors which rely on TLS
handshake fingerprinting~\cite{Frolov2019a} to identify \tool.

As a concrete example, consider the setting where a client $C$ streams music
from an audio streaming server $S$. The two parties would like to use this
channel to send data from $S$ to $C$. \tool assumes a trusted setup phase where
both $C$ and $S$ agree on a symmetric key and playlist of songs; that is, $C$
knows \emph{a priori} some subset of the songs $S$ will stream. On launch, $S$
starts the audio streaming application (e.g., Icecast) with \tool, which
intercepts outgoing traffic produced by the application and replaces the audio
data with a pointer to where in the playlist the given audio data corresponds.
On $C$'s end, \tool intercepts incoming traffic to $C$'s listening application
(e.g., VLC) and replaces the data with the actual audio data (which $C$ knows,
as this info was pre-shared), before passing on the data to the listening
application.

Because network reads/writes originate within the (unmodified) application, their lengths and
behavioral characteristics---modulo slight timing differences introduced by the
processing required by \tool---\emph{exactly} match that of the application
running without \tool. The \tool framework also provides a generic signaling technique to
allow clients and servers to covertly mutually authenticate each other. Because the
server runs an unmodified application binary, it could even be providing a legitimate service
(such as a public audio streaming channel in the above example). Normal clients
can successfully connect to the \tool-enabled server as usual, without detecting
anything about its circumvention capability.

\newcommand{\rotation}[1]{#1}
\begin{table*}[t]
    \small
    \centering
    \begin{threeparttable}
    \begin{tabular}{@{}ccccccc@{}}
                      &                   & \textbf{Unmodified} & \textbf{Standard} & \textbf{Does Not Require}\\
      \textbf{Scheme} & \textbf{Approach} & \textbf{Binary} & \textbf{Input} & \textbf{Encryption} & \textbf{Flexible} & \textbf{Goodput}\\
      \midrule
      FTE\cite{Dyer2013a} & Mimicry & N/A & & $\checkmark$ & $\checkmark$ & 1.9--42 mbps\tnote{$\ast$} \\
      DeltaShaper\cite{Barradas2017a} & Tunneling & $\checkmark$ & & & & 2.56 kbps \\
      Freewave\cite{Houmansadr2013a} & Tunneling & $\checkmark$ & & & & 19 kbps \\
      Castle\cite{Hahn2016a} & Tunneling & $\checkmark$ & $\checkmark$ & & & 190 bps \\
      Rook\cite{Vines2015a} & Tunneling & $\checkmark$ & $\checkmark$ & $\checkmark$ & & 26-34 bps \\
      Protozoa\cite{Barradas2020} & Tunneling & & $\checkmark$ & & & 160--1400 kbps\\
      \midrule
      \tool (audio streaming) & \multirow{2}{*}{Tunneling} & \multirow{2}{*}{$\checkmark$} & \multirow{2}{*}{$\checkmark$} & & \multirow{2}{*}{$\checkmark$} & 145 kbps\tnote{$\ast\ast$} \\
      \tool (web browsing) &&&&&& 8 mbps\tnote{$\dagger$}
    \end{tabular}
    {\scriptsize
      \begin{tablenotes}
      \item[$\ast$] This range corresponds to an HTTP format on the low-end, and
          an SSH format on the high-end.
      \item[$\ast\ast$] When streaming an audio file encoded at 148 kbps.
      \item[$\dagger$] When downloading a video with bandwidth capped at 8 mbps.
      \end{tablenotes}
    }
    \end{threeparttable}
    \caption{Comparison of several look-like-something link obfuscation schemes
      versus \tool. ``Unmodified Binary'' denotes those schemes that run an
      unmodified implementation of the target protocol under-the-hood,
      ``Standard Input'' denotes those schemes that run on input that matches
      the expected input of the implementation, ``Does Not Require Encryption''
      denotes those schemes that do not rely on encryption for undetectability,
      ``Flexible'' denotes those schemes which provide frameworks for supporting
      various applications/protocols, and ``Goodput'' denotes the covert
      throughput of the scheme. }\label{table:link-obf}
\end{table*}

\autoref{table:link-obf} provides a comparison of \tool to several mimicry and
tunneling approaches (see also our discussion of related work in
\autoref{sec:relatedwork}). While \tool is not the first CRS to use standard
input to drive the channel, it is the first to provide a flexible framework
while achieving significantly higher goodput than prior work.

\tool, however, is not a panacea. It specifically relies on TLS and the fact
that TLS is not being man-in-the-middled by a censor. In environments where TLS
is expressly forbidden or actively man-in-the-middled (which occurs from time to
time~\cite{Kazakhstan}), \tool may be detectable. Also, like most CRSs, \tool
does not address the channel setup phase, the phase most often attacked by
censors~\cite{Tschantz2016a}. However, despite these drawbacks, \tool offers a
flexible framework for building circumvention channels, one which generalizes
prior approaches and which can be adjusted, by varying the model or application,
to the characteristics of the network environment in which it is being deployed.

\subsection{Our Contributions}

To summarize, we make the following contributions:

\begin{itemize}
\item We introduce \tool, an open-source framework for censorship circumvention
    which embeds data in TLS-protected traffic generated by an \emph{unmodified}
    application binary. \tool is designed to make it easy to spin-up new
    instantiations for different applications and protocols. While the high
    level idea of \tool is relatively straightforward, realizing an
    implementation is quite complicated due to the need to minimize the effect
    \tool has on packet timings alongside avoiding subtle attack vectors; see
    \autoref{sec:arch} for the architecture description and
    \autoref{sec:implementation} for implementation details.

\item We describe two instantiations of \tool (\autoref{sec:instantiations}):
    one for audio streaming and one for web browsing. In the audio streaming
    case, \tool is able to replace all of an audio stream with arbitrary
    data---when streaming an Ogg-Vorbis file with a bitrate of 148
    kilobit/second this corresponds to a 148 kilobit/second channel. In the web
    browsing case, \tool is able to replace all content transmitted via
    \texttt{HTTP} including HTML, CSS, image, audio, and video files.

\item We provide a security analysis (\autoref{sec:security}) and evaluation
    (\autoref{sec:evaluation}) of \tool against both passive and active
    adversaries.
\end{itemize}
Because the \tool framework is extensible to new protocols and new applications, we believe that
its deployment could help enable censorship circumvention providers to evolve
more quickly in response to developments of a censor's capability.

\section{Architecture}\label{sec:arch}

\tool provides a bidirectional\footnote{The bidirectionality is dependent on the
  application and network protocol used; for example, our audio streaming
  instantiation only achieves a unidirectional channel.} channel-based
censorship circumvention framework for TLS-protected channels. The framework
needs to be instantiated for specific applications/protocols. In this work we
demonstrate two such instantiations: (1) audio streaming and (2) web browsing.
We assume the censor monitors the network traffic between the two communicating
parties and can use both passive and active attacks to identify the channel. We
also assume a trusted setup phase where the communicating parties agree on some
shared information: a symmetric key and a \emph{traffic model} which encodes the
particular plaintext data to replace (cf.\ \autoref{sec:model}).

\begin{figure}[t]
    \centering
    \def\svgwidth{\columnwidth}
    \scalebox{1}{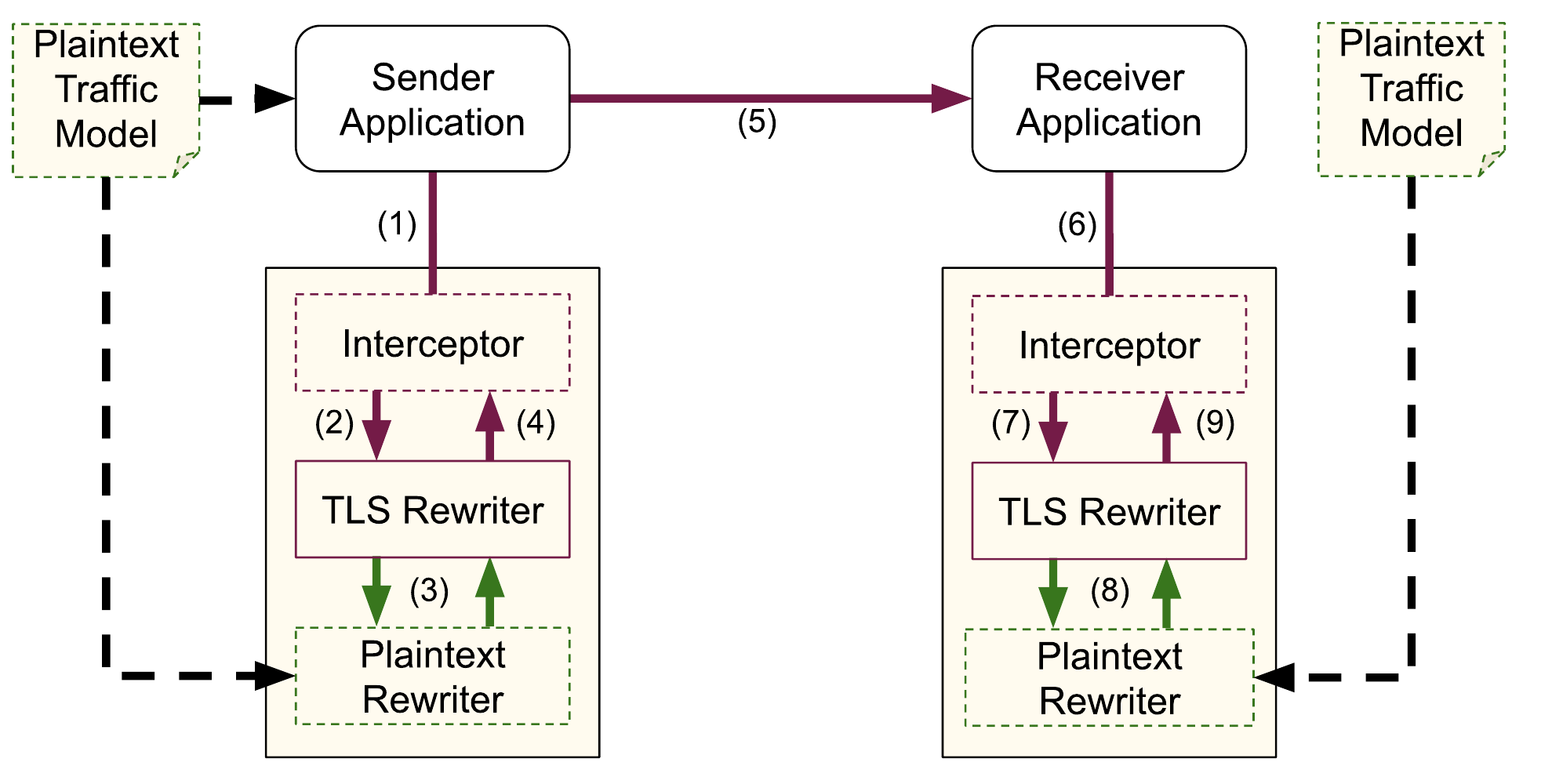}
    \vspace{-4ex}
    \caption{The \tool architecture. \textcolor{black}{Yellow} denotes \tool
      components, \textcolor{mymagenta}{red} denotes TLS-encrypted data, and
      \textcolor{mygreen}{green} denotes plaintext data. Boxes with dashed lines
      denote instantiation-specific components of \tool.}
    \label{fig:arch}
\end{figure}

\autoref{fig:arch} shows the overall \tool architecture. \tool sits between an
application and the network, intercepting outgoing/incoming TLS streams (Steps~1
and~6). The intercepted stream is then fed to a \emph{TLS rewriter} (Step~2),
which extracts the underlying plaintext of the TLS stream. For outbound traffic,
the plaintext is fed to a protocol-specific \emph{plaintext rewriter} that
replaces the plaintext with a pointer to the appropriate location in the traffic
model and fills in the leftover bytes with any data to send (Step~3). For
inbound traffic, the plaintext is again fed to a protocol-specific plaintext
rewriter that extracts the data and replaces the model pointer with
the pointed-to data (Step~8). The TLS rewriter then re-encrypts the
(transformed) plaintext data before feeding it back to the calling application
(Steps~4 and~9).

In what follows we walk through this architecture in more detail, discussing the
relevant implementation considerations along the way.

\subsection{Traffic Models}\label{sec:model}

\tool makes use of \emph{traffic models} that capture some subset of the
expected plaintext network traffic between the communicating parties, and \tool
assumes that the communicating parties have access to compatible models.
While the traffic model
structure is specific to a particular \tool instantiation, \emph{within a given
  \tool instantiation} the particular traffic model may differ between each pair
of communicating parties. For example, client $C_1$ talking to audio streaming
server $S$ may use a different traffic model than client $C_2$ talking to the
\emph{same} server~$S$. We discuss the traffic model structures for our
instantiations in \autoref{sec:instantiations}.

Importantly, the traffic model need not be a model of the \emph{entire}
interaction between the parties. This allows parties to communicate $N$ bytes of
data without needing the model to be of size $O(N)$. In addition, for
bidirectional instantiations of \tool, the traffic model could even be
\emph{learned} by the client, who could then update the server on the traffic
model to use. For example, for web browsing---assuming a base traffic model that
allows the parties the communicate---the client could collect a set of assets
available on the server to use as its traffic model and inform the server on
which assets to use going forward.

Additionally, the traffic model need not be static. For example, in the audio
streaming setting, the server could dynamically generate audio from a seed and
send that seed along with covert data to the client. The \tool client could then
replicate the dynamically-generated music that the server is sending to all the
non-\tool clients. For web browsing, the server could be running a blog in which
the articles are automatically generated from some seed (enabling them to be
replaced with covert data for a \tool client), while comments (which can be
posted by arbitrary users) can be sent through unmodified.

\subsection{Potential Deployment Scenarios}\label{sec:deployment}

Due to \tool's use of both a shared key and traffic model between the
communicating parties, we believe \tool's ideal deployment scenario is one in
which a small trusted set of clients (such as a select set of journalists)
are aware that a given server is \tool-enabled. Recall that the \tool-enabled
server functions exactly as a server would without \tool running, and thus this
server could provide a service to the public at large. For example,
the server could be a programming blog, providing the set of trusted clients a
reasonable ``alibi'' for accessing the server.

\subsection{Intercepting TLS Data}

\tool needs to intercept outgoing TLS data (Step~1) in order to rewrite the
underlying plaintext before sending it to the receiver, and needs to intercept
incoming TLS data (Step~6) to extract the covert data before sending the
(original) plaintext on to the application. In \tool, we use dynamic linker
features to manipulate network traffic by intercepting calls to \texttt{libc}
system call wrappers. This approach has two distinct advantages over alternative
approaches: (1) since we are directly running an unmodified version of the
application, the network traffic characteristics exactly match those of the
application (besides slight timing differences), and (2) the approach is more
amenable to adding support for additional applications (or additional application versions)
since we can largely treat the application as a black box.

\subsubsection{Implementing Dynamic Library Injection}\label{sec:implementing-dynamic-library-injection}
On Linux\footnote{While our primary target operating system is Linux, \tool
  works on macOS using \texttt{DYLD\_INSERT\_LIBRARIES} (and other features of
  the macOS dynamic linker) instead of \texttt{LD\_PRELOAD}. } \tool takes
advantage of the \texttt{LD\_PRELOAD} option to \texttt{ld.so} to perform
dynamic library injection. The dynamic linker causes calls to \texttt{read()},
\texttt{write()}, \texttt{sendmsg()}, \texttt{writev()}, among others, to be
captured by \tool instead of performing their usual action inside the C standard
library. \tool's injection library is tuned to the particular protocol to
specify (1)~which network connections to intercept (e.g., based on IP address or
port number), and (2)~which plaintext rewriter to use for the particular
protocol/application.

This approach does have several subtle considerations that complicate the
implementation, which we discuss below.

\paragraph{Performance considerations.}
Because \tool performs in-band network traffic rewriting, it operates on the
``hot path'', and thus any delay imposed by \tool's processing may be directly
visible to a censor monitoring the connection. Thus, it is vital that \tool is
as efficient as possible. As a result, \tool's rewriter code is designed to be
low-latency. We achieve this primarily by avoiding memory allocation alongside
implementing a high-performance logging library (see
\autoref{sec:implementation}), among other standard techniques. We discuss
specific performance numbers in \autoref{sec:eval:microbenchmarks}.

\paragraph{Recursive calls.}
\tool may invoke \texttt{libc} functions as part of its operation. If such a
call occurs within an \emph{intercepted} \texttt{libc} function this could cause
an infinite loop.
\tool mitigates this by maintaining a flag in thread-local storage
to see whether control has already entered an injected function call. If so, then the
\texttt{libc} routine that \tool replaced is transparently called instead.

\paragraph{Signal safety.}
Several functions that \tool intercepts are considered \emph{signal-safe} by the
POSIX standard. As a result, an application might call any of these functions
from inside a signal-handler. \tool mitigates this issue via the same recursive
call mechanism described above. That being said, \tool is not perfectly
signal-safe---more extensive testing and implementation work is necessary to
ensure full signal safety.

\paragraph{Limitations of dynamic library injection.}
Because we use dynamic library injection, \tool does not work on applications
that do not use dynamic library calls to perform network operations (such as an
application written in Go)\footnote{We note that \tool \emph{still} works even
  if the TLS library is statically linked, as long as the TLS library supports
  extracting the TLS key material through the \texttt{SSLKEYLOGFILE} environment
  variable.}. In addition, because we only intercept POSIX (and Linux) network
APIs, we restrict ourselves to Unix-like operating systems; in particular, we do
not have Windows support for \tool. However, this could potentially be added
using DLL injection techniques; we leave this to future work.

\subsection{Extracting TLS Key Material}\label{sec:key-material}

In order for \tool to manipulate TLS data it must first learn the TLS key
material. It does so by taking advantage of debugging features available in most
modern TLS libraries.

\paragraph{\texttt{SSLKEYLOGFILE}.} When working with an application using GnuTLS,
NSS\footnote{Mozilla's TLS library, used in Firefox and Thunderbird, among other software.}, or
Rustls\footnote{A TLS library written in Rust:
  \url{https://github.com/ctz/rustls}}, \tool constructs a named pipe and
passes it to the application using the \texttt{SSLKEYLOGFILE} environment
variable. The application sends a serialized form of the TLS master secret
to \tool which can use it for further processing.

\paragraph{OpenSSL.} OpenSSL does not support the \texttt{SSLKEYLOGFILE}
environment variable. Thus, when working with an application that dynamically
links to OpenSSL, \tool uses \texttt{LD\_PRELOAD} to inject a shim over the
\texttt{SSL\_new()} function that configures a callback to receive the TLS key
material. For applications that \emph{statically} link to OpenSSL, we rely on
the application itself to support \texttt{SSLKEYLOGFILE}; this is the case for
many applications, including curl, among many others.

\medskip

Because \tool treats the application's TLS library as a gray-box---that is, the
only requirement beyond using \texttt{libc} system call wrappers is that the TLS
library supports dumping the TLS key material in some way---\tool has a single
TLS rewriter codebase that works with OpenSSL, GnuTLS, NSS, and Rustls. Since
\tool is very weakly-coupled to the application's TLS library, it makes it easy
to extend support to additional applications, as well as additional TLS
libraries. As an example, no code changes were required to get \tool working for
Rustls once we implemented GnuTLS support.

A significant benefit of extracting TLS key material from the library itself
is that \tool \emph{does not modify the
  TLS handshake}. This prevents a whole class of attacks that censors commonly
employ to detect CRSs\cite{Frolov2019a}.
One downside however is that \tool cannot make any active changes to the TLS
traffic until the key information has been emitted. Fortunately, every TLS
library that we looked at releases the TLS master secret by the time a TLS
Application Record is sent or received, which is sufficient for \tool's needs.

\subsection{Processing Intercepted TLS Data}\label{sec:arch:processing}

Once \tool has intercepted the TLS data, the next steps are to: (1) decrypt the
data, (2) rewrite the resulting plaintext, and (3) re-encrypt the
plaintext to either send over the wire or return to the application. We describe each
of these steps in turn.

\subsubsection{Decrypting TLS Data}

\tool decrypts incoming and outgoing TLS data (Steps~2 and~7) identically. How
decryption works depends on the particular TLS version and cipher suite used. In
particular, \tool currently only supports TLS 1.2 and stream cipher suites (see
\autoref{sec:tls13} and \autoref{sec:appendix-non-aead} for a discussion on how
we can support TLS 1.3 and non-stream cipher suites, respectively, although we
leave the implementation to future work).
To decrypt, \tool scans the intercepted TLS data for Application Data records,
ignoring other record types\footnote{ \tool also looks for Alert records. If an
  Alert record is observed, then \tool transparently passes traffic to the
  application without modifying it. }. Once it has found an Application Data
record, it reads the explicit nonce for the record (if there is one\footnote{ In
  TLS 1.2, the ChaCha20-Poly1305 cipher takes the approach that is standard in
  TLS 1.3 of having no explicit nonce sent over the wire. }). Armed with the
explicit nonce, \tool performs an unauthenticated decryption of the bytes over
the wire. As these bytes are decrypted, they are sent to the plaintext rewriter
for processing. After the payload has been processed, \tool reads the (original)
MAC of the incoming record, and checks that it is correct. If it is, \tool
generates a new MAC for the rewritten record, and if not, \tool generates an
invalid MAC\@. While the above gives the high-level idea, we discuss some
subtleties with this approach in \autoref{sec:arch:one-byte}.

\subsubsection{Rewriting the Plaintext}\label{sec:compressing}

Given the extracted plaintext data, \tool either rewrites the plaintext to make
room for user-provided data (Step~3) or extracts the user-provided data and
rewrites the plaintext to recover the original data (Step~8). Rewritten bytes
are then forwarded on for re-encryption.

How rewriting is performed is protocol (and possibly application) specific and
must be designed on a per-protocol basis. This is the key point at which \tool
is configurable. We have implemented two instantiations of \tool---audio
streaming and web browsing---which we discuss in \autoref{sec:instantiations}.

\subsubsection{Re-encryption}\label{sec:re-encrypt}

The final step is to re-encrypt the plaintext before sending it either over
the wire (Step~4) or to the application itself (Step~9). For the former case, we
could simply re-encrypt using the extracted TLS master secret; however, this
leaves open the possibility that a censor that man-in-the-middles the TLS
connection could extract the user data. We thus re-encrypt using a key $k'$
derived from the TLS master secret $mk$ and the pre-shared key $k$. That is,
$k' \leftarrow \KDF(mk \| k)$,
where \KDF is a key derivation function (BLAKE3 in our case). Besides this
change, re-encryption operates the same for Steps 4 and 9.

\subsubsection{Handling Partial Reads and Writes}\label{sec:arch:one-byte}

In order to be as faithful to the application's behavior as possible, \tool
rewrites TLS data immediately upon intercepting a system call. If a system call
returns an error (such as \texttt{EWOULDBLOCK}), then \tool forwards that
response on to the caller.\footnote{An alternative approach would be to perform
  multiple, e.g., \texttt{read()}s upon intercepting a \texttt{read()}. However,
  such an approach would potentially alter the TCP flow control in a sufficient
  way to be identifiable to a censor.}
The immediate rewriting, however, results in several implementation
complications, which we elaborate on below.

\begin{figure}[t]
    \centering
    \def\svgwidth{\columnwidth}
    \scalebox{1}{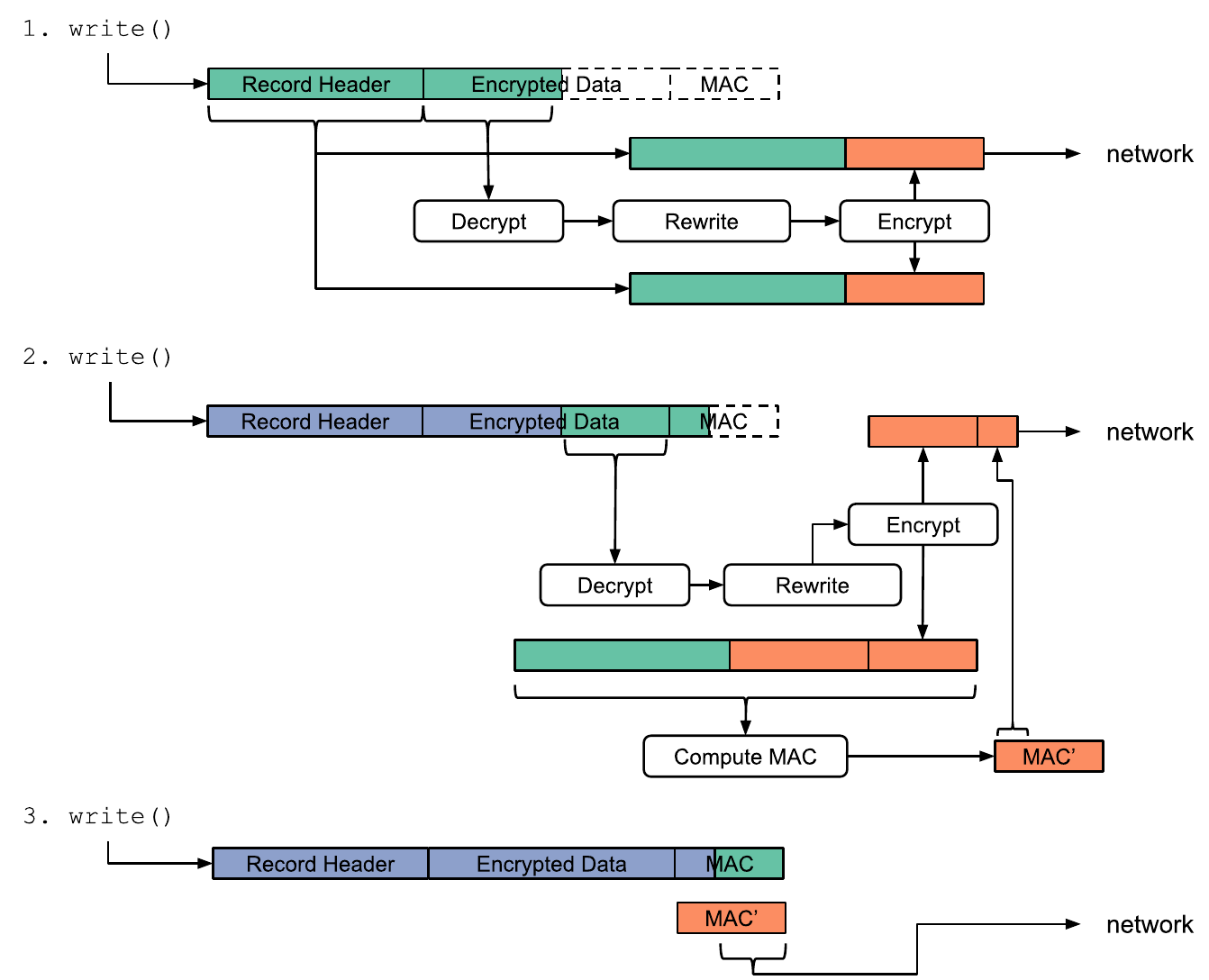}
    \caption{Processing outgoing TLS records. We consider a scenario where it
      takes three calls to the \texttt{write()} function for the application to
      write the full TLS record. \textcolor{mygreen2}{Green} denotes data
      written during a given \texttt{write()} call,
      \textcolor{mypurple2}{purple} denotes prior written data, and
      \textcolor{myorange2}{orange} denotes data computed by \tool. }
    \label{fig:processing-outgoing-data}
\end{figure}

\paragraph{Handling partial writes.}
For performance purposes, TLS libraries optimistically try to \texttt{write()}
as much data as possible. In practice, this means that \tool gets to see at
least one full TLS record in a single intercepted \texttt{write()}. However, if
the application's TLS library attempts to write more bytes than there is room
for in the kernel's buffer, then the kernel reports that only a partial write
occurred. \tool handles this by performing unauthenticated decryption until the
MAC is received. \autoref{fig:processing-outgoing-data} provides an illustrated
example, where it takes three \texttt{write()}s to emit a complete TLS record.

\paragraph{Handling partial reads.}
Handling \texttt{read()}s is more complicated as \emph{the number of bytes that
  \texttt{read()} returns may depend on censor-controlled network conditions}.
As a result, unlike with \texttt{write()}s, where we know that we should see
whole chunks at a time, with \texttt{read()}s a censor could manipulate the TCP
connection such that each successful \texttt{read()} \emph{only yields one
  byte}. In order to cope with this, we designed \tool to be able to
decide what byte to return to the application given only a single incoming byte
alongside any previously observed traffic. In particular, when processing one
byte at a time \tool does not necessarily have access to the given TLS record's
MAC (that is, it may not be contained in the data acquired for the particular
\texttt{read()} function call made by the application), and so it cannot
authenticate the TLS record until all bytes of the TLS record have been
received. However, \tool must provide \emph{something} to the application on
each \texttt{read()} call, and this something must be the re-encrypted plaintext
data if the MAC is indeed correct. \tool addresses this conundrum by
\emph{assuming} that the TLS record is valid, up until the last byte of the
incoming MAC, providing an invalid value for the last MAC byte if it turns out
that the incoming MAC was incorrect.

\begin{figure}[t]
    \centering
    \def\svgwidth{\columnwidth}
    \scalebox{1}{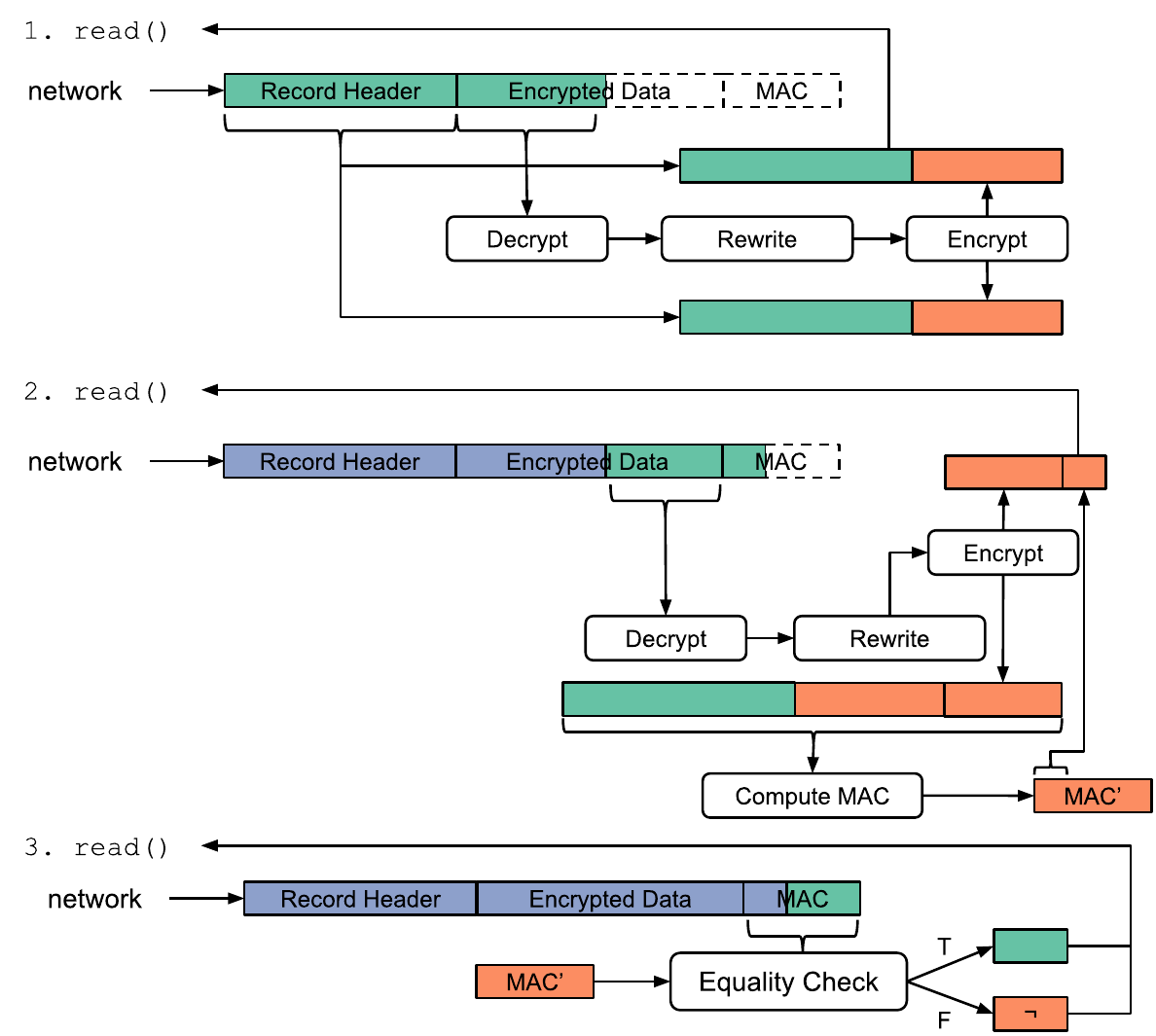}
    \caption{Processing incoming TLS records. We consider a scenario where it
      takes three calls to the \texttt{read()} function for the application to
      read the full TLS record. \textcolor{mygreen2}{Green} denotes data
      read during a given \texttt{read()} call, \textcolor{mypurple2}{purple}
      denotes prior read data, and \textcolor{myorange2}{orange} denotes data
      computed by \tool.}
    \label{fig:processing-incoming-data}
\end{figure}

\autoref{fig:processing-incoming-data} provides an illustrated example of how
\tool handles this. In the figure, the application makes three calls to the underlying
\texttt{read()} function to read the full TLS record. In the first
\texttt{read()}, \tool has not yet received the MAC so cannot actually validate
that the incoming TLS record is valid. It thus assumes it is, sending back the
re-encrypted plaintext data to the application. In the second \texttt{read()}, \tool
receives a portion of the MAC\@. Again, it cannot assume the MAC is correct, but
must provide the plaintext data alongside a portion of the MAC to the application.
In this case, it computes the expected MAC (MAC$'$ in the figure) and passes the
requisite portion of MAC$'$ to the application. Finally, in the third \texttt{read()},
\tool receives the full MAC\@. It does an equality check between this MAC and
its precomputed one: if these MACs are equal then the TLS record is valid, and
\tool sends the rest of the MAC on to the application. Otherwise, it sends the inverse
of the last byte of the MAC to force the application to receive an invalid MAC (which is
what the application would have received in the case where \tool was not used).

\subsection{Signaling}\label{sec:covert-signaling}

While the above steps allow parties to communicate using \tool, an important
step is for the parties to signal that they want to send/receive data in the
first place. \tool's signaling protocol allows the client and server to
authenticate to each other, and is designed to be secure even against active
probes made by the censor. We assume a secret key $k$ has been pre-shared
between the client and server, and use that---in conjunction with the TLS master
secret---to derive a server key $k_S$ and client key.

\subsubsection{How the Client Authenticates the Server}
We re-use the existing certificate mechanism in TLS for the client to authenticate
the server. \tool clients are provisioned with a pinned public key certificate
which is validated against the signature that the server sends in its Server Key
Exchange TLS Handshake record. If the signature does not match, \tool enters a
transparent pass-through state and makes no modification to the traffic.

\subsubsection{How the Server Authenticates the Client}
The main challenge with signaling is for the server to authenticate the
client. \tool's protocol has two settings: (1) one in which it assumes
that the server waits for a TLS Application Data record from the client before
it sends any Application Data itself (as is the case in HTTP and other
protocols), and (2) one in which it does not make this assumption.

\paragraph{Setting \#1.} When \tool intercepts the client's first Application
Data record, it leaves the plaintext untouched but replaces the MAC $T$ with $T
\oplus k_C$. Because the client has already verified the server's certificate as
part of the key exchange, the censor is unable to distinguish between $T$ and
$T \oplus k_c$.

On the server, \tool looks for the incoming client-sent Application Data record.
\tool then checks to see whether $T$ or $T \oplus k_C$ is a valid
MAC for the given record. If $T$ is a valid MAC then the server assumes
it is dealing with a non-\tool client and
enters a transparent pass-through state in which it performs no traffic modification.
If $T \oplus k_C$ is a valid MAC, then signaling has succeeded and the
rewriting stages can proceed as normal.
If neither $T$ nor $T \oplus k_C$ is a valid MAC, then \tool
passes an intentionally invalid MAC to the application and enters its transparent pass-through
state. This case may occur if the censor has tampered with the
connection, and by passing an invalid MAC to the application, \tool causes it to respond as it
would ordinarily to an invalid MAC\@.

\paragraph{Setting \#2.}
If the client does \emph{not} always send an Application Data record before the
server, then \tool proceeds as follows. \tool on the server starts by transparently
passing-through all outgoing Application Data records. When \tool on the client
receives these records, it also transparently passes them on to its application.

\tool on the client performs the same operation as in Setting \#1 on the first
client-sent Application Data record. The client has now successfully completed
its \emph{outgoing} signaling efforts, and can now freely perform its
normal plaintext rewriting and re-encryption processes on its outgoing traffic.

Because TLS (and TCP) are \emph{full-duplex} protocols, there is no ordering
relationship between client-to-server messages and server-to-client messages.
The client can immediately proceed with its normal outgoing rewriting processes
because the ordering constraints of TCP and TLS ensure that the server sees the
Application Data message with the mangled MAC before it sees any messages sent
after that. However, when a message comes in from the server, the client does not
know whether that message was sent before or after the server saw the client's
initial signaling message (in the form of the mangled MAC). As a result,
the client does not know which key (namely, the standard TLS master secret or
the derived re-encryption key) to use to decrypt the message. In addition, the
client does not know whether to attempt to rewrite the message.
To reiterate, the problem is the following: the client knows that the server is
a \tool-server, and it has told the server that it is a \tool-client, but
because incoming and outgoing messages have no ordering relationship, the client
does not know whether the server \emph{knows} that the client is a \tool-client.

We solve this problem by having the server \emph{acknowledge} that it received
the client's initial signal. By having the server signal on its outgoing
half of the duplex connection, any subsequent messages that it sends will arrive
after its acknowledgement message. The server sends its acknowledgment message
by replacing the MAC $T$ on an outgoing Application Data record with $T \oplus
k_S$. After sending this message, the server can start its normal \tool
operations.
The client scans incoming Application Data records and performs the same
check from above to find an Application Data record where the MAC is $T \oplus
k_S$. After observing that message, the client is free to start rewriting
incoming traffic from the server.

\subsubsection{Security of Signaling}
Several CRSs~\cite{Bocovich2016a,Ellard2015a} use a signaling technique based on
Telex~\cite{Wustrow2011a} which modifies the Client Random field of the TLS
Client Hello message. While this change is indistinguishable to a censor, we do
not use this technique because it would require us to re-implement many more
pieces of TLS, and it would not work with our method of using TLS libraries'
debugging features to extract TLS key material. In addition, Telex's signaling
scheme does not offer forward secrecy: a censor can record network traffic and
then, if at any point in the future they compromise the server, they would be
able to go back through the recorded traffic and determine which connections used
signaling.

In contrast, \tool's signaling scheme inherits the forward secrecy of
TLS\@: because the key material that \tool uses to perform signaling is
based on the ephemeral key of the TLS connection, any future compromise of the
server would not reveal which connections had signaling. As a result, \tool's
shared covert signaling secret has the same security properties of Telex's public
key: any client with the key can authenticate itself to the server, but the key
does not allow any client (except for the sever) to identify which clients are
using the key.

\section{\tool Instantiations}\label{sec:instantiations}

We have implemented two instantiations of \tool: one for audio streaming and one
for web browsing. We describe each in turn.

\subsection{Audio Streaming}\label{sec:audio-streaming}

This instantiation supports Ogg Vorbis audio streaming traffic generated by an
Icecast instance, with the client running a media player such as VLC\footnote{We
  have in addition validated that \tool works for several other media players,
  including Audacious, Rhythmbox, etc.}. The traffic model in this case is a
single Ogg Vorbis file containing a concatenation of audio files.

Our rewriter works specifically for Ogg Vorbis traffic. Vorbis is a free and
patent-free audio coding format (similar to MP3), and Ogg provides a container
format for transmitting Vorbis streams. Icecast streams audio data to the client
in an \texttt{HTTP/1.0} response, which does not terminate. Ogg data itself is
broken up into \emph{pages}, each of which starts with an Ogg page
header.

When the rewriter encounters an Ogg page, it determines whether the page is a
candidate to be rewritten. A page is ``rewriteable'' if its body can be found in
the source audio (i.e., the traffic model). Because an Ogg page might not fit
entirely in a single TLS record, the rewriter sometimes has to decide whether a
page is rewriteable before seeing it in its entirety. To get around this, the
rewriter searches for audio data prefixed by what it has learned is in the body.
It then uses the CRC32 checksum present in the original Ogg page to determine
whether its guess of the audio data was correct. If the rewriter is unable to
find a match, then it passes the page through
unmodified.

If the rewriter does decide to rewrite an Ogg page, it replaces the Version
field, which is normally a `0', with `*'\footnote{The choice of `*' as the
  ``version'' is arbitrary.}. This Version field signals to the receiver that it
should attempt to rewrite the page. Next, the rewriter replaces the Bitstream
Serial Number with the byte offset in the original audio data to which the data
in the page corresponds. With the page header modified, we can replace the
\emph{entire} audio data component with covert data.

To rewrite an Ogg page on the receiver's side, we first check whether the page
corresponds to covert data by checking that the Version field in the page header
is the magic number `*'. If so, we extract the data and then replace it with the
actual audio data using the location specified in the Bitstream Serial Number.

\subsection{Web Browsing}\label{sec:web-browsing}

This instantiation handles web browsing between a Firefox client and an Apache web
server. We consider a traffic model in which the communicating parties share a
directory of shared assets, such as HTML, images, video files, etc., and
currently only support a unidirectional covert channel between the server and
client.

Our rewriter works by parsing the \texttt{HTTP} request made by the client and
stores the \texttt{HTTP} version, verb, request URI, and headers. For example, a
request to \texttt{https://example.com/dir/index.html} might have a version of
\texttt{HTTP/1.1}, a \texttt{GET} verb, \texttt{/dir/index.html} as the request
URI, and header values for fields such as %
\texttt{Host}, \texttt{User-Agent}, and \texttt{Cookie}. Similarly, when the
server receives the \texttt{HTTP} request, its rewriter parses and stores the
request information. The server's rewriter waits until an \texttt{HTTP} response
is sent in reply. The rewriter parses the response to extract the status code
and headers. If the status code indicates success and the request URI matches a
shared asset, the body of the \texttt{HTTP} response is overwritten with covert
data. In addition, to indicate to the client that rewriting has occurred, the
third byte of the
\texttt{\textbackslash{}r\textbackslash{}n\textbackslash{}r\textbackslash{}n}
bytes between the response header and body is rewritten to \texttt{0xff}. When
the client receives the response, its rewriter parses the response. If the
\texttt{0xff} byte is present, the rewriter extracts the covert data and
replaces it with the shared asset data.

\texttt{HTTP} allows partial downloads of files, which is often used for
streaming audio or video files. Our \texttt{HTTP} rewriter supports this
functionality by first checking for a \texttt{206 Partial Content} status code.
It then checks for \texttt{Content-Range} headers in the \texttt{HTTP} response
and rewrites the shared asset with the appropriate position offset and length
based on values in the range header.

\section{Implementation}\label{sec:implementation}

We have implemented \tool alongside rewriters for audio streaming and web
browsing. \tool is implemented in Rust.

\paragraph{Code organization.}
\tool is comprised of several Rust crates that correspond to the components
depicted in \autoref{fig:arch}:
\begin{itemize}
\item \texttt{injection} contains the core code and traits for injecting code
    into a shared library. A rewriter for \tool needs to provide implementations
    of the associated traits for the particular application being injected.

\item \texttt{tlsRewriter} contains code for rewriting the TLS records, and
    handles the decryption and re-encryption required. We have tested the
    rewriter with the following TLS libraries: OpenSSL, GnuTLS, and Rustls\@.

\item \texttt{rewriter} contains the traits for implementing protocol-specific
    (plaintext) rewriters. An instantiation of \tool needs to provide
    implementations of these traits.

\end{itemize}
Because \tool must contend with partial reads (cf.\
\autoref{sec:arch:one-byte}), it can be tedious to manually write a state
machine to perform byte manipulations. To remedy this, the \texttt{rewriter}
and \texttt{tlsRewriter} components are written as
coroutines. Coding in this style makes the rewriter implementations smaller and
easier to develop.

For our audio streaming rewriter, we implemented the rewriter described in
\autoref{sec:audio-streaming} and implemented wrapper code for injecting \tool
into VLC and Icecast. This wrapper code is reusable across multiple multimedia
clients; for example, the wrapper code works for Audacious, Rhythmbox, MPlayer,
and mpv, among others, without requiring a single line of code to be changed
from the original VLC implementation.

Our web browsing rewriter proceeded similarly: we implemented the rewriter
described in \autoref{sec:web-browsing} and implemented wrapper code for
injecting \tool into Firefox and the Apache Web Server. We have additionally
tested the Firefox injector on curl.

\paragraph{High speed logging.}
We developed a highly-performant logging library called Stallone to facilitate
debugging \tool both during implementation and for any potential future
deployment. Due to the careful performance considerations required, we could not
use existing logging libraries, as those add overheads of hundreds of
microseconds per log entry, which would add noticeable delay to a running \tool
instance. We thus designed Stallone from scratch, taking inspiration from the
NanoLog library~\cite{Yang2018}. Compared to NanoLog, Stallone does not rely on
the CPU's timestamp counter, which might not be stable or valid in cloud
environments or in any situation where the user does not know what exact CPU
model they are working with\cite{DontRDTSC}. In addition, Stallone uses stable
identifiers for log record types and stores the mapping between log record
identifiers and log record metadata (such as the message and line number) in a
special section of the binary, eliminating the need for this information to be
dumped online.
Stallone is written in Rust and is capable of
logging messages at an overhead of around 10 nanoseconds, and as such may be of
independent interest.

\section{Security Analysis}\label{sec:security}

In this section we discuss the security of \tool versus a censor that controls
all network traffic between the communicating parties, and either passively
monitors the network or actively manipulates, blocks, or injects packets. Due to
the heavy systems engineering and subtle implementation details required in
building \tool---alongside a lack of security definitions within the field of
censorship circumvention---we forgo a formal (i.e., ``provable security'')
treatment of \tool. Instead, given the relative simplicity of the cryptography
inside \tool, we focus more closely on the practical security of the
implementation (and the timing channel that it produces).

\paragraph{Identifying the signaling protocol.} \tool's signaling
protocol (cf.\ \autoref{sec:covert-signaling}) replaces the original MAC of a
TLS record with a one-time-pad of the MAC and a key derived from the TLS master
secret and the pre-shared secret. Because the master secret is chosen
pseudorandomly for each connection, and because the censor does not know the
pre-shared secret, the new MAC is indistinguishable from the original to a
censor.

However, \tool's signaling protocol does leave open the possibility of a
timing channel resulting from the need to compute the modified MAC and check
equality when an invalid MAC is encountered. We minimize this channel by
precomputing the \KDF as soon as the TLS master secret is known, reducing the
online cost to a single XOR operation.

\paragraph{Manipulating the TLS channel.} \tool alters the TLS channel by
replacing the plaintext data in a given TLS record.
This replacement is indistinguishable from standard application traffic,
assuming the security of TLS.
However, due to restrictions on reading from the network (cf.\
\autoref{sec:arch:one-byte}), \tool currently requires the use of a
\emph{stream} cipher suite. Thus, an active censor could force
a particular cipher suite to be used, one that is not supported by \tool. Thus,
\tool \emph{only} operates for specific supported cipher suites, and otherwise
operates in pass-through mode. This however leaves open the possibility of a
denial-of-service attack where a censor actively enforces that only
non-stream cipher modes are negotiated. We view such an attack as highly
unlikely, given that 81\% of TLS connections use stream cipher
suites~\cite{icsi-notary}.
However, even in this case we can resort to
supporting non-streaming modes as discussed in \autoref{sec:appendix-non-aead}.

A sufficiently powerful censor may be able to man-in-the-middle the TLS
connection and thus recover the covert data. Such attacks are not
unrealistic~\cite{Kazakhstan}. While we cannot prevent such a censor from
\emph{identifying} that \tool is in use, we prevent the censor from acquiring
the covert data by re-encrypting it using a different key than the TLS master
secret, as specified in \autoref{sec:re-encrypt}.

\paragraph{Manipulating the application itself.}
A censor could try to use traffic manipulation or injection to force \tool
to enter an invalid state, producing behavior that is distinguishable from what
the underlying application would have done. We carefully designed \tool such that
whenever it reaches a failure mode it reverts to pass-through mode such that
any observer sees the underlying application behavior directly.

\paragraph{Identifying timing differences.} The main difference between running
the application with or without \tool is the timing differences introduced by \tool. We
discuss the effects these timing differences have on classifying \tool for
audio streaming and web browsing in \autoref{sec:evaluation}.

\paragraph{Identifying plaintext traffic model differences.} A censor may try to
identify \tool by identifying differences between a particular traffic model and
the baseline behavior of the network environment. As an example, if an audio
streaming service streams the same song over and over the traffic pattern may
differ sufficiently from other audio streaming services found on the network.
Note that this attack is \emph{external} to whether \tool is deployed. That is,
if the user's behavior varies significantly from behavior in the baseline
network environment, a (sufficiently powerful) censor could detect this
\emph{whether or not \tool was running at all}\footnote{Whether such an attack
  is feasible in practice depends heavily on the censor and what their false
  positive threshold is.}. Thus, it is important to choose an appropriate
traffic model instantiation for the particular deployment environment of \tool,
and this choice is one that needs to be made with the particular deployment
environment in mind (e.g., the expected audio from a stream in Country~A may
differ from that in Country~B).

\paragraph{Mimicking a client.} A censor can try to determine a \tool server by
acting as a client. Assuming the censor does not have the required shared key to
allow it to signal the server, the probability it successfully guesses the
modified MAC and hence passes the signaling protocol is negligible.

\paragraph{Mimicking a server.} A censor could also mimic a server, flagging
any client that connects and produces a TLS record with an invalid MAC\@. \tool
thwarts this attack by verifying the public-key signature in the TLS connection
against a pinned public-key. If this verification fails, then \tool enters a
transparent pass-through mode, and the connection appears as normal to the server.

\section{Evaluation}\label{sec:evaluation}

There are several avenues in which we evaluate \tool: goodput and detectability.
As discussed in \autoref{sec:security}, the ability for a censor to identify
\tool depends in part on any delay introduced by the tool over the baseline
performance of the application. Thus, we focus our detectability evaluation on
(1) producing microbenchmarks for the delay introduced by our two instantiations
of \tool, and (2) building classifiers for \tool under various network latency
settings to investigate whether a passive censor could detect \tool.

\subsection{Goodput}\label{sec:eval:goodput}

Because \tool tunnels data through existing channels, the goodput of \tool
closely matches the throughput of the cover channel. In particular, for audio
streaming we can replace 98\% of cover data. Thus, when streaming an audio file
encoded at $X$ kbps ($X = 148$ or $X = 160$ is standard), we
achieve a goodput of $.98 \cdot X$. For web browsing the computation is slightly
more complicated, as the precentage of data we can replace depends on the size
of the cover asset. For example, if the asset is a blank HTML page we would
achieve a very low goodput as there is no cover data to replace. However, for the
``real-world'' assets we have tested against (everything from single HTML pages
to video files) we have found that we can replace 62--99\% of cover
data.

\subsection{Microbenchmarks}\label{sec:eval:microbenchmarks}

As discussed in \autoref{sec:security}, \tool introduces timing delays due to
the processing required to rewrite TLS records and perform plaintext rewriting.
To measure this delay, we ran \tool on a standard laptop (Intel Core i7-6820HQ @
2.7~GHz) for both our audio streaming and web browsing rewriters, tracking the
cost of each rewrite operation for the sender and receiver. Each rewrite
consists of decrypting the TLS data (encrypted under the \texttt{AES128-GCM-SHA256}
or \texttt{AES256-GCM-SHA384} cipher suites), rewriting the plaintext, and
re-encrypting---that is, a rewrite
consists of all the processing done by \tool upon intercepting a \texttt{read()}
or \texttt{write()} from the underlying application.

\paragraph{Audio streaming.}
We gathered data while streaming a 10 second Ogg Vorbis audio file encoded at a
bitrate of 148 kbps. For the sender (i.e., Icecast), we see an average delay of
$122 \mu s$. The delay seen on the receiver depends on the particular client
application we are running; for example, for VLC we see an average delay of $36
\mu s$ and for MPlayer we see an average delay of $20 \mu s$. The additional
delay imposed by the sender is largely due to (1) the CRC computation required
when replacing the plaintext Ogg data, and (2) the computation of the GCM tag
required when re-encrypting the plaintext.

\paragraph{Web browsing.} We gathered data for two scenarios: using curl to
download a video file and using Firefox to browse several links on a website
containing a small subset of Wikipedia. For the sender (i.e., Apache), we see an
average delay across both scenarios of roughly $89 \mu s$. For curl we see an
average delay of $90 \mu s$, and for Firefox we see an average delay of $216 \mu
s$. The reason we see a higher delay than audio streaming is that the web
browsing rewriter needs to store \texttt{HTTP} requests and thus requires
allocations.

\subsection{Timing Analysis}\label{sec:eval:timing}

The introduced delays have security implications, as a sufficiently powerful
censor may be able to classify \tool-enabled traffic due to these delays. To
determine the effect of these timing differences on the ability to classify
\tool we ran several experiments on both our audio streaming and web browsing
instantiations. For all of our experiments, we generated 130 \texttt{pcap}
traces\footnote{ We generated packet captures on an {Intel Xeon Silver 4114 CPU
    @ 2.20GHz} with 40 cores and 512 GB of memory. Doing so enabled us to
  generate packet captures more quickly, by running multiple trials in parallel.
  We (informally) verified that running parallel trials did not impact our
  results by comparing the results against a small number of non-parallel runs.
} between two Ubuntu 18.04 docker containers with and without \tool enabled, using
\texttt{tc} to control the average latency and its standard deviation in our
simulated network. We generated traces for latencies between 0 ms (the ``ideal''
scenario) and 30 ms (the average latency in the United States\footnote{According
  to \url{http://ipnetwork.bgtmo.ip.att.net/pws/averages.html} as of January 28,
  2020.}). We then built classifiers to try to distinguish the \tool-enabled
versus -disabled traffic, using \texttt{tcptrace}~\cite{tcptrace} to extract TCP
statistics to train on. Our classifiers use random forests due to the success
similar classifiers have had on distinguishing prior censorship circumvention
systems~\cite{Barradas2018a}. For each scenario we trained classifiers using
10-fold stratified cross-validation using Scikit-learn~\cite{scikit-learn}.

Note that all of these experiments occurred in an idealized setting with no
additional network traffic, and thus represent a \emph{best case scenario for a
  censor}. In a real-world deployment we would expect that successfully applying
such a classifier to be much more difficult. We additionally experimented with
running our experiments alongside a number of ``dummy'' clients, which act
exactly as the normal client except their network data is not analyzed by the
classifier. We found---as expected---that this \emph{decreases} the classifier's
accuracy\footnote{As an example, for VLC with zero latency and four ``dummy''
  clients, we achieve a classifier accuracy of only 66\%, versus 84\% when a
  single client is used.}, and thus to model the best case scenario for a censor we consider the
single-client setting.

\paragraph{Audio streaming.}

\begin{table*}[t]
    \footnotesize
    \centering
    \begin{subtable}[b]{0.45\textwidth}
        \centering
        \begin{tabular}{@{}cccc@{}}
          \textbf{Latency (ms)} & \textbf{Accuracy} & \textbf{Precision} & \textbf{Recall}\\
          \midrule
          \midrule
          0 $\pm$ 0 & 0.84 $\pm$ 0.07 & 0.87 $\pm$ 0.09 & 0.80 $\pm$ 0.11 \\ %
          \midrule
          5 $\pm$ 1 & 0.72 $\pm$ 0.08 & 0.76 $\pm$ 0.10 & 0.66 $\pm$ 0.13 \\ %
          5 $\pm$ 3 & 0.63 $\pm$ 0.09 & 0.67 $\pm$ 0.11 & 0.55 $\pm$ 0.14 \\ %
          \midrule
          10 $\pm$ 1 & 0.67 $\pm$ 0.09 & 0.71 $\pm$ 0.12 & 0.59 $\pm$ 0.13 \\ %
          10 $\pm$ 3 & 0.67 $\pm$ 0.09 & 0.70 $\pm$ 0.11 & 0.61 $\pm$ 0.14 \\ %
          10 $\pm$ 5 & 0.59 $\pm$ 0.09 & 0.61 $\pm$ 0.11 & 0.51 $\pm$ 0.14 \\ %
          \midrule
          30 $\pm$ 1 & 0.64 $\pm$ 0.09 & 0.67 $\pm$ 0.11 & 0.57 $\pm$ 0.14 \\ %
          30 $\pm$ 3 & 0.56 $\pm$ 0.09 & 0.57 $\pm$ 0.12 & 0.47 $\pm$ 0.15 \\ %
          30 $\pm$ 5 & 0.57 $\pm$ 0.09 & 0.58 $\pm$ 0.12 & 0.49 $\pm$ 0.14 \\ %
          30 $\pm$ 10 & 0.50 $\pm$ 0.09 & 0.50 $\pm$ 0.12 & 0.41 $\pm$ 0.14 \\ %
        \end{tabular}
        \caption{VLC}
    \end{subtable}
    \hfill
    \begin{subtable}[b]{0.45\textwidth}
        \centering
        \begin{tabular}{@{}cccc@{}}
          \textbf{Latency (ms)} & \textbf{Accuracy} & \textbf{Precision} & \textbf{Recall}\\
          \midrule
          \midrule
          0 $\pm$ 0 & 0.68 $\pm$ 0.09 & 0.72 $\pm$ 0.12 & 0.60 $\pm$ 0.13 \\ %
          \midrule
          5 $\pm$ 1 & 0.50 $\pm$ 0.10 & 0.50 $\pm$ 0.12 & 0.41 $\pm$ 0.14 \\ %
          5 $\pm$ 3 & 0.51 $\pm$ 0.09 & 0.51 $\pm$ 0.12 & 0.41 $\pm$ 0.14 \\ %
          \midrule
          10 $\pm$ 1 & 0.55 $\pm$ 0.10 & 0.56 $\pm$ 0.12 & 0.47 $\pm$ 0.15 \\ %
          10 $\pm$ 3 & 0.53 $\pm$ 0.09 & 0.54 $\pm$ 0.12 & 0.45 $\pm$ 0.14 \\ %
          10 $\pm$ 5 & 0.52 $\pm$ 0.09 & 0.53 $\pm$ 0.12 & 0.42 $\pm$ 0.14 \\ %
          \midrule
          30 $\pm$ 1 & 0.53 $\pm$ 0.10 & 0.54 $\pm$ 0.13 & 0.44 $\pm$ 0.14 \\ %
          30 $\pm$ 3 & 0.49 $\pm$ 0.10 & 0.49 $\pm$ 0.12 & 0.40 $\pm$ 0.14 \\ %
          30 $\pm$ 5 & 0.50 $\pm$ 0.09 & 0.50 $\pm$ 0.12 & 0.41 $\pm$ 0.13 \\ %
          30 $\pm$ 10 & 0.49 $\pm$ 0.09 & 0.48 $\pm$ 0.12 & 0.39 $\pm$ 0.14 \\ %
        \end{tabular}
        \caption{MPlayer}
    \end{subtable}
    \par\bigskip
    \begin{subtable}[b]{0.45\textwidth}
        \centering
        \begin{tabular}{@{}cccc@{}}
          \textbf{Latency (ms)} & \textbf{Accuracy} & \textbf{Precision} & \textbf{Recall}\\
          \midrule
          \midrule
          0 $\pm$ 0 & 0.82 $\pm$ 0.05 & 0.85 $\pm$ 0.07 & 0.78 $\pm$ 0.08 \\ %
          \midrule
          5 $\pm$ 1 & 0.73 $\pm$ 0.06 & 0.75 $\pm$ 0.07 & 0.71 $\pm$ 0.09 \\ %
          5 $\pm$ 3 & 0.68 $\pm$ 0.06 & 0.70 $\pm$ 0.07 & 0.63 $\pm$ 0.09 \\ %
          \midrule
          10 $\pm$ 1 & 0.68 $\pm$ 0.06 & 0.70 $\pm$ 0.07 & 0.63 $\pm$ 0.10 \\ %
          10 $\pm$ 3 & 0.59 $\pm$ 0.07 & 0.61 $\pm$ 0.08 & 0.53 $\pm$ 0.10 \\ %
          10 $\pm$ 5 & 0.63 $\pm$ 0.07 & 0.65 $\pm$ 0.08 & 0.56 $\pm$ 0.10 \\ %
          \midrule
          30 $\pm$ 1 & 0.65 $\pm$ 0.06 & 0.68 $\pm$ 0.08 & 0.59 $\pm$ 0.10 \\ %
          30 $\pm$ 3 & 0.56 $\pm$ 0.06 & 0.57 $\pm$ 0.08 & 0.48 $\pm$ 0.10 \\ %
          30 $\pm$ 5 & 0.59 $\pm$ 0.07 & 0.61 $\pm$ 0.08 & 0.52 $\pm$ 0.10 \\ %
          30 $\pm$ 10 & 0.56 $\pm$ 0.07 & 0.58 $\pm$ 0.08 & 0.49 $\pm$ 0.10 \\ %
        \end{tabular}
        \caption{Audacious}
    \end{subtable}
    \hfill
    \begin{subtable}[b]{0.45\textwidth}
        \centering
        \begin{tabular}{@{}cccc@{}}
          \textbf{Latency (ms)} & \textbf{Accuracy} & \textbf{Precision} & \textbf{Recall}\\
          \midrule
          \midrule
          0 $\pm$ 0 & 0.66 $\pm$ 0.09 & 0.69 $\pm$ 0.11 & 0.61 $\pm$ 0.13 \\ %
          \midrule
          5 $\pm$ 1 & 0.53 $\pm$ 0.09 & 0.54 $\pm$ 0.12 & 0.44 $\pm$ 0.14 \\ %
          5 $\pm$ 3 & 0.57 $\pm$ 0.09 & 0.58 $\pm$ 0.12 & 0.48 $\pm$ 0.14 \\ %
          \midrule
          10 $\pm$ 1 & 0.55 $\pm$ 0.09 & 0.57 $\pm$ 0.12 & 0.46 $\pm$ 0.14 \\ %
          10 $\pm$ 3 & 0.49 $\pm$ 0.09 & 0.48 $\pm$ 0.13 & 0.38 $\pm$ 0.14 \\ %
          10 $\pm$ 5 & 0.53 $\pm$ 0.09 & 0.54 $\pm$ 0.13 & 0.43 $\pm$ 0.14 \\ %
          \midrule
          30 $\pm$ 1 & 0.53 $\pm$ 0.09 & 0.53 $\pm$ 0.12 & 0.43 $\pm$ 0.14 \\ %
          30 $\pm$ 3 & 0.53 $\pm$ 0.09 & 0.54 $\pm$ 0.12 & 0.44 $\pm$ 0.14 \\ %
          30 $\pm$ 5 & 0.52 $\pm$ 0.10 & 0.53 $\pm$ 0.13 & 0.42 $\pm$ 0.15 \\ %
          30 $\pm$ 10 & 0.50 $\pm$ 0.10 & 0.49 $\pm$ 0.13 & 0.40 $\pm$ 0.14 \\ %
        \end{tabular}
        \caption{mpv}
    \end{subtable}
    \caption{Accuracy, precision, and recall of classifying \tool-generated
      traffic versus baseline for various latency settings against various media
      players (VLC, MPlayer, Audacious, and mpv). Values are given in ``mean
      $\pm$ standard deviation'' format. }\label{table:classification}
\end{table*}

For audio streaming we investigated the potential to identify \tool running
across four different media players: VLC, MPlayer, Audacious, and mpv.
Each trace comprised of a client connecting to an Icecast server, streaming a 10
second song, and then exiting.
\autoref{table:classification} presents the accuracy, precision, and recall of
our classifier for different latencies against these different media players.
For each scenario we trained \num{1000} classifiers, with the presented results
being the average and standard deviation of these classifiers.

We find at the extreme end---where there is zero latency in the simulated
network---the classifier is able to distinguish \tool traffic across the various
media players with between 66\% and 84\% accuracy, with the key features being
the average TCP window advertisement seen and data transmit time. This suggests
that even the slight delay introduced by \tool is enough to affect some network
statistics (albeit in an unrealistic network setting).
However, as we increase the realism of the network (by increasing the average
latency as well as the standard deviation) we see the accuracy of the classifier
quickly drop to a point where it is essentially no better than random guessing.
This makes sense given that the delays introduced by \tool become part of the
noise of the network latency.

Another interesting feature of \autoref{table:classification} is that the
classifer accuracy varies depending on the media player. This suggests (perhaps
not surprisingly) that different media players present different ``network
footprints''. To validate this, we additionally ran our classifier to see if we
could distinguish two different media players, both with \tool disabled. We
found that regardless of which media players we compared against, we achieved a
99--100\% accuracy for all latency and standard deviation
settings.

\paragraph{Web browsing.}
For web browsing we investigated the potential to identify \tool using two
different clients: curl and Firefox. For curl, each trace comprised of
downloading a 13.6 MB video and then exiting. For Firefox, each trace comprised of
a Selenium script accessing three different web pages scraped from Wikipedia,
sleeping three seconds between each web page access.
The assets for the three web pages totaled 8.9 MB and included HTML, javascript, image, and CSS files.
As with the audio streaming
case, \autoref{table:classification-web} presents the accuracy, precision, and
recall of our classifier across different latencies.

While the accuracies for web browsing tend to be higher than in the audio
streaming case, this makes sense given the larger average delay introduced by
\tool. However, we reiterate that these results are for an \emph{ideal} setting
for the censor and the accuracies are still sufficiently low given the base rate
fallacy.

\begin{table*}[t]
    \footnotesize
    \centering
    \begin{subtable}[b]{0.45\textwidth}
        \centering
        \begin{tabular}{@{}ccccc@{}}
          \textbf{Latency (ms)} & \textbf{Accuracy} & \textbf{Precision} & \textbf{Recall}\\
          \midrule\midrule
          0 $\pm$ 0 & 0.66 $\pm$ 0.01 & 0.68 $\pm$ 0.01 & 0.61 $\pm$ 0.01 \\ %
          \midrule
          5 $\pm$ 1 & 0.69 $\pm$ 0.01 & 0.71 $\pm$ 0.01 & 0.64 $\pm$ 0.01 \\ %
          5 $\pm$ 3 & 0.69 $\pm$ 0.01 & 0.71 $\pm$ 0.01 & 0.64 $\pm$ 0.01 \\ %
          \midrule
          10 $\pm$ 1 & 0.66 $\pm$ 0.01 & 0.68 $\pm$ 0.01 & 0.61 $\pm$ 0.01 \\ %
          10 $\pm$ 3 & 0.66 $\pm$ 0.01 & 0.68 $\pm$ 0.01 & 0.60 $\pm$ 0.01 \\ %
          10 $\pm$ 5 & 0.65 $\pm$ 0.01 & 0.67 $\pm$ 0.01 & 0.58 $\pm$ 0.01 \\ %
          \midrule
          30 $\pm$ 1 & 0.69 $\pm$ 0.01 & 0.71 $\pm$ 0.01 & 0.66 $\pm$ 0.02 \\ %
          30 $\pm$ 3 & 0.67 $\pm$ 0.01 & 0.69 $\pm$ 0.01 & 0.62 $\pm$ 0.01 \\ %
          30 $\pm$ 5 & 0.62 $\pm$ 0.01 & 0.63 $\pm$ 0.01 & 0.55 $\pm$ 0.02 \\ %
          30 $\pm$ 10 & 0.57 $\pm$ 0.01 & 0.59 $\pm$ 0.01 & 0.49 $\pm$ 0.02 \\ %
        \end{tabular}
        \caption{Firefox}
    \end{subtable}
    \hfill
    \begin{subtable}[b]{0.45\textwidth}
        \begin{tabular}{@{}ccccc@{}}
          \textbf{Latency (ms)} & \textbf{Accuracy} & \textbf{Precision} & \textbf{Recall}\\
          \midrule\midrule
          0 $\pm$ 0 & 0.96 $\pm$ 0.04 & 0.97 $\pm$ 0.05 & 0.95 $\pm$ 0.06 \\ %
          \midrule
          5 $\pm$ 1 & 0.71 $\pm$ 0.08 & 0.74 $\pm$ 0.11 & 0.65 $\pm$ 0.13 \\ %
          5 $\pm$ 3 & 0.66 $\pm$ 0.09 & 0.69 $\pm$ 0.12 & 0.58 $\pm$ 0.14 \\ %
          \midrule
          10 $\pm$ 1 & 0.79 $\pm$ 0.08 & 0.81 $\pm$ 0.09 & 0.77 $\pm$ 0.12 \\ %
          10 $\pm$ 3 & 0.70 $\pm$ 0.08 & 0.73 $\pm$ 0.10 & 0.64 $\pm$ 0.13 \\ %
          10 $\pm$ 5 & 0.63 $\pm$ 0.09 & 0.67 $\pm$ 0.12 & 0.56 $\pm$ 0.14 \\ %
          \midrule
          30 $\pm$ 1 & 0.86 $\pm$ 0.07 & 0.90 $\pm$ 0.08 & 0.82 $\pm$ 0.11 \\ %
          30 $\pm$ 3 & 0.62 $\pm$ 0.09 & 0.65 $\pm$ 0.12 & 0.55 $\pm$ 0.14 \\ %
          30 $\pm$ 5 & 0.62 $\pm$ 0.09 & 0.65 $\pm$ 0.12 & 0.55 $\pm$ 0.14 \\ %
          30 $\pm$ 10 & 0.67 $\pm$ 0.08 & 0.71 $\pm$ 0.11 & 0.58 $\pm$ 0.13 \\ %
        \end{tabular}
        \caption{curl}
    \end{subtable}
    \caption{Accuracy, precision, and recall of classifying \tool-generated
      traffic versus baseline for various latency settings against various web
      clients (curl and Firefox). Values are given in ``mean
      $\pm$ standard deviation'' format. }\label{table:classification-web}
\end{table*}

\section{Related Work}\label{sec:relatedwork}

The literature is rich with different approaches to building censorship
resistant systems (CRSs); we refer the reader to existing systematization of
knowledge papers~\cite{Khattak2016a,Tschantz2016a} for a more thorough overview
of the field than what we can provide here.

A CRS can be viewed as comprising two key components: \emph{communication
  establishment} and \emph{conversation}. \tool addresses the second,
which is where most of the academic literature has focused~\cite[\S
5.5]{Khattak2016a}. In particular, \tool corresponds to an ``access-centric''
scheme using the terminology of Khattak et al.~\cite{Khattak2016a}. We thus
focus on such schemes in this section. Access-centric schemes can be subdivided
into four\footnote{Khattak et al.~\cite{Khattak2016a} differentiate between
  \emph{tunneling} approaches and \emph{covert channel} approaches whereas we
  view these as the same, since any covert channel approach necessarily needs to
  ``tunnel'' its traffic through some existing application.} main categories,
which we discuss in turn.

\paragraph{Mimicry.} These approaches send data by mimicking some cover
protocol. A representative example is \emph{format-transforming
  encryption}~\cite{Dyer2013a} and its variants~\cite{Luchaup2014a,Dyer2015a},
which operate by mapping ciphertexts to regular expressions or context-free
grammars that can encode, e.g., common network protocols like HTTP\@. The
well-known ``Parrot is Dead'' paper~\cite{Houmansadr2013b} argues that such
approaches are doomed to fail due to the difficulty of accurately mimicking a
given protocol, although as discussed below (and in \autoref{sec:introduction}) even
tunneling approaches suffer the same challenges.

\paragraph{Tunneling.} These approaches try to avoid the ``weaknesses'' of the
mimicry approach by running the actual application under-the-hood. Such
approaches include Freewave~\cite{Houmansadr2013a},
DeltaShaper~\cite{Barradas2017a}, and Castle~\cite{Hahn2016a}. However, as
several researchers have
shown~\cite{Geddes2013a,Wang2015a,SkypeEncryptionIsNotEnough}, even these
approaches are susceptible to distinguishing attacks due the protocol
distribution differences between the circumvention system itself and the
underlying application when run on its own. This weakness appears inherent due
to the inability to perfectly mimic the real world application behavior, or let
alone know what such a ``real world distribution'' is in the first place. \tool
aims to minimize this gap by having such real world application behavior be a
parameter specified by the user of the tool.

Concurrently with this work, Barradas et al.~\cite{Barradas2020} introduced
Protozoa, a tunneling approach which uses WebRTC as its communication medium.
Protozoa shares several similarities to \tool, in that it uses a form of
rewriting to replace WebRTC traffic with user data. However, Protozoa is
specific for WebRTC and requires modifications to the application source code,
reducing the flexibility of the tool as application versions change, an attack
vector exploited in practice~\cite{Frolov2019a}. It also does not replace the
original video on the receiver side, potentially leaving the approach open to
traffic analysis attacks.

\paragraph{Traffic manipulation.} These approaches manipulate traffic to
circumvent known censors. Recent approaches, such as Geneva~\cite{Bock2019a},
have proven successful at circumventing existing nation-state censors in several
countries. However, the security model is fundamentally different (and weaker)
than the one considered by both \tool and tools in the mimicry and tunneling
space: traffic manipulation approaches generally assume a weak censor that
monitors traffic using a firewall or deep packet inspection device, whereas
\tool considers a potentially active censor that can apply more powerful
capabilities. (Whether this more powerful censor is a realistic threat in
practice is an orthogonal question.)

\paragraph{Destination obfuscation.} These approaches, which include Tor and
refraction networking protocols~\cite{Wustrow2011a,Nasr2017a,Bocovich2016a},
focus on hiding the destination website from a censor, and borrow from the
mimicry and tunneling literature in how they obfuscate the channel itself (e.g.,
Tor uses a ``pluggable transport'' infrastructure for link obfuscation).

\paragraph{Other related work.}
Several CRSs either require a specific version of an application (such as
meek~\cite{Fifield2015a}) or otherwise need to mimic the TLS handshake in some
way. However, Frolov and Wustrow~\cite{Frolov2019a} showed that this mimickry is
often easily identifiable due to cleartext header information sent in the
initial \emph{Client Hello} message of a TLS connection---that is, this
information must exactly match what an innocuous (and popular) application would
produce. With this in mind, the authors introduce a tool, uTLS, for
automatically mimicking existing TLS implementations.

\tool avoids the need for a tool like uTLS by running the (unmodified)
application under-the-hood and leaving the TLS handshake untouched. As long as
the underlying protocol used by the application remains unchanged between
versions, the application can be updated without affecting \tool. In particular,
unlike tools like meek~\cite{Fifield2015a}, \tool does not need to come bundled
with a particular version of an application.

\section*{Acknowledgments}
This material is based upon work supported by the United States Air Force and
DARPA under Contract No. FA8750-19-C-0085. Any opinions, findings and
conclusions or recommendations expressed in this material are those of the
author(s) and do not necessarily reflect the views of the United States Air
Force and DARPA. Distribution Statement ``A'' (Approved for Public Release,
Distribution Unlimited).

\bibliographystyle{plain}
\bibliography{citations,censorbib}

\appendix

\section{Supporting TLS 1.3}\label{sec:tls13}

One nice feature of TLS 1.2 is that handshake records are distinct from
application records, and are distinguished by early bytes in the record
header. However, this is not the case for TLS 1.3: handshakes may occur at any
time during a given connection and are distinguished by the last encrypted byte
of the encrypted payload. As a result, when operating on incoming TLS 1.3 records,
\tool does not know whether the record should be rewritten or not.

Our proposed solution to this problem is to add functionality to the sender's
plaintext rewriter to let it rewrite the first byte of a TLS 1.3 handshake
record (which contains the TLS record handshake type) into a form that the
receiver's rewriter can distinguish from the rewriting of the first plaintext
byte of an Application Data record. As an example, the rewriter could set the
high-order bit of the first byte of the record in an HTTP request to denote that
it is Application Data and not a Handshake record. \tool could then use this
information to determine whether to proceed with rewriting.

\section{Supporting CBC-mode Ciphers}\label{sec:appendix-non-aead}

While \tool's current implementation only supports stream ciphers, it is
possible for \tool to intercept TLS traffic encrypted with a CBC-mode cipher and
still operate under the restriction that incoming traffic can be processed one
byte at-a-time. We leave the implementation of the below approach as future
work.

To avoid the numerous number of attacks on CBC-mode, modern TLS libraries use a
randomly generated initialization vector (IV) for each TLS record. \tool can
take advantage of this fact as follows. For outgoing traffic, \tool can replace
the TLS record IV with the encryption (under a stream cipher, with an IV from
the sequence number) of the first block of plaintext. It can then proceed in
this fashion, replacing each subsequent block of ciphertext with the
stream-cipher-encryption of the next block of plaintext. The last block of CBC
encrypted ciphertext can be replaced with random noise. The MAC can be handled
as in the case for stream ciphers.

On the incoming side, because the incoming plaintext is encrypted with a stream
cipher, \tool can decrypt it one byte at-a-time. To re-encrypt the traffic with
a CBC-mode cipher for the application, \tool can pick a new random IV to encrypt
the block with, and it emit this random IV\@. Even with the one byte at-a-time
requirement, by the time \tool emits any encrypted bytes of the plaintext it
would have already observed a full block of plaintext, enabling it to generate
the encrypted bytes. \tool can then rewrite the outgoing MAC, in the same manner
as it would for stream ciphers, to generate a MAC that matches the ciphertext
that it just outputted (if the incoming MAC is valid).

\end{document}